\documentstyle{mn}

\newif\ifAMStwofonts

\def\ueber#1#2{{\setbox0=\hbox{$#1$}%
  \setbox1=\hbox to\wd0{\hss$\scriptscriptstyle #2$\hss}%
  \offinterlineskip
  \vbox{\box1\kern0.4mm\box0}}{}}
\def \mathrm{\rm}

\def\Omegam{\Omega_{\rm m}}

\def\lsim{\stackrel{<}{{}_\sim}}
\def\gsim{\stackrel{>}{{}_\sim}}

\def\e{{\mathrm{e}}}

\def\d{{\mathrm{d}}}

\catcode`@=11
\def\gsim{\mathrel{\mathpalette\@versim>}}
\def\@versim#1#2{\lower0.2ex\vbox{\baselineskip\z@skip\lineskip\z@skip
       \lineskiplimit\z@\ialign{$\m@th#1\hfil##$\crcr#2\crcr\sim\crcr}}}
\catcode`@=12
\catcode`@=11
\def\lsim{\mathrel{\mathpalette\@versim<}}
\def\@versim#1#2{\lower0.2ex\vbox{\baselineskip\z@skip\lineskip\z@skip
       \lineskiplimit\z@\ialign{$\m@th#1\hfil##$\crcr#2\crcr\sim\crcr}}}
\catcode`@=12
\def\etal{{et al.~}}

\def\km{${\rm km\; s}^{-1}$}
\newcommand\be{\begin{equation}}
\newcommand\ee{\end{equation}}
\newcommand{\ba}{\begin{array}}
\newcommand{\ea}{\end{array}}

\ifoldfss
  \ifCUPmtlplainloaded \else
    \NewTextAlphabet{textbfit} {cmbxti10} {}
    \NewTextAlphabet{textbfss} {cmssbx10} {}
    \NewMathAlphabet{mathbfit} {cmbxti10} {} 
    \NewMathAlphabet{mathbfss} {cmssbx10} {} 
  \fi
  \ifAMStwofonts
    \ifCUPmtlplainloaded \else
      \NewSymbolFont{upmath} {eurm10}
      \NewSymbolFont{AMSa} {msam10}
      \NewMathSymbol{\upi}     {0}{upmath}{19}
      \NewMathSymbol{\umu}     {0}{upmath}{16}
      \NewMathSymbol{\upartial}{0}{upmath}{40}
      \NewMathSymbol{\leqslant}{3}{AMSa}{36}
      \NewMathSymbol{\geqslant}{3}{AMSa}{3E}

      \let\leq=\leqslant 
      \let\geq=\geqslant 
    \fi
  \fi
\fi 

\ifnfssone
  \newmathalphabet{\mathit}
  \addtoversion{normal}{\mathit}{cmr}{m}{it}
  \addtoversion{bold}{\mathit}{cmr}{bx}{it}
  \newmathalphabet{\mathbfit} 
  \addtoversion{normal}{\mathbfit}{cmr}{bx}{it}
  \addtoversion{bold}{\mathbfit}{cmr}{bx}{it}
  \newmathalphabet{\mathbfss} 
  \addtoversion{normal}{\mathbfss}{cmss}{bx}{n}
  \addtoversion{bold}{\mathbfss}{cmss}{bx}{n}
  \ifAMStwofonts
    \ifCUPmtlplainloaded \else
      \UseAMStwoboldmath
      \makeatletter
      \new@mathgroup\upmath@group
      \define@mathgroup\mv@normal\upmath@group{eur}{m}{n}
      \define@mathgroup\mv@bold\upmath@group{eur}{b}{n}
      \edef\UPM{\hexnumber\upmath@group}
      \new@mathgroup\amsa@group
      \define@mathgroup\mv@normal\amsa@group{msa}{m}{n}
      \define@mathgroup\mv@bold\amsa@group{msa}{m}{n}
      \edef\AMSa{\hexnumber\amsa@group}
      \makeatother
      \mathchardef\upi="0\UPM19
      \mathchardef\umu="0\UPM16
      \mathchardef\upartial="0\UPM40
      \mathchardef\leqslant="3\AMSa36
      \mathchardef\geqslant="3\AMSa3E

      \let\leq=\leqslant 
      \let\geq=\geqslant 
    \fi
  \fi
\fi 

\ifnfsstwo
  \DeclareMathAlphabet{\mathbfit}{OT1}{cmr}{bx}{it}
  \SetMathAlphabet\mathbfit{bold}{OT1}{cmr}{bx}{it}
  \DeclareMathAlphabet{\mathbfss}{OT1}{cmss}{bx}{n}
  \SetMathAlphabet\mathbfss{bold}{OT1}{cmss}{bx}{n}
  \ifAMStwofonts
    \ifCUPmtlplainloaded \else
      \DeclareSymbolFont{UPM}{U}{eur}{m}{n}
      \SetSymbolFont{UPM}{bold}{U}{eur}{b}{n}
      \DeclareSymbolFont{AMSa}{U}{msa}{m}{n}
      \DeclareMathSymbol{\upi}{0}{UPM}{"19}
      \DeclareMathSymbol{\umu}{0}{UPM}{"16}
      \DeclareMathSymbol{\upartial}{0}{UPM}{"40}
      \DeclareMathSymbol{\leqslant}{3}{AMSa}{"36}
      \DeclareMathSymbol{\geqslant}{3}{AMSa}{"3E}

      \let\leq=\leqslant 
      \let\geq=\geqslant 
    \fi
  \fi
\fi 

\ifCUPmtlplainloaded \else
  \ifAMStwofonts \else 
    \def\upi{\pi}
    \def\umu{\mu}
    \def\upartial{\partial}
  \fi
\fi

\title{Overconstrained dynamics in galaxy redshift surveys}
\author[M. Susperregi]
       {Mikel Susperregi\\
Fisika Teorikoaren eta Zientzi Historiaren Saila, Zientzi Fakultatea, 
Euskal Herriko Unibertsitatea, PO Box 644, 48080 Bilbao, Spain\\
Email: {\sf wtpsuxxm@ehu.es}}

\date{Version of \today}

\pagerange{0--0}
\pubyear{0000}

\begin{document}

\maketitle

\label{firstpage}

\begin{abstract}
The least-action principle (LAP) method is used on four galaxy 
redshift surveys to measure the density parameter $\Omegam$ and 
the matter and galaxy-galaxy power spectra. The datasets are 
PSC$z$, ORS, Mark III and SFI. The LAP method is applied on the surveys 
simultaneously, resulting in an overconstrained dynamical system that 
describes the cosmic overdensities and velocity flows. The system 
is solved by relaxing the constraint that each survey 
imposes upon the cosmic fields. A least-squares optimization of the 
errors that arise in the process yields the cosmic fields and the value of $\Omegam$ 
that is the best fit to the ensemble of datasets. The analysis has been carried out with a 
high-resolution Gaussian smoothing of 500 \km and over a spherical selected volume of 
radius 9,000 \km. We have assigned a weight to each survey, depending on 
their density of sampling, and this parameter determines their relative 
influence in limiting the domain of the overall solution. The influence of each 
survey on the final value of $\Omegam$, the cosmographical features of the 
cosmic fields and the power spectra largely depends on the distribution function of 
the errors in the relaxation of the constraints. We find that PSC$z$ and Mark III    
are closer to the final solution than ORS and SFI. The likelihood analysis yields  
$\Omegam= 0.37\pm 0.01$ to 1$\sigma$ level. PSC$z$ and SFI are the closest to this 
value, whereas ORS and Mark III predict a somewhat lower $\Omegam$. The model of bias employed 
is a scale-dependent one, and we retain up to 42 bias coefficients 
$b_{rl}$ in the spherical harmonics formalism. The predicted power spectra are estimated 
in the range of wavenumbers $0.02\,h\,{\rm Mpc}^{-1}\lsim k\lsim 0.49\,h\,{\rm Mpc}^{-1}$,  
and we compare these results with measurements recently reported in the literature.  
\end{abstract}

\begin{keywords}
cosmology: theory -- large-scale structure of Universe  
-- cosmic flows -- galaxies: distances, velocities and redshifts
\end{keywords}

\section{Introduction}

Galaxy redshift surveys are elements of prime importance in the 
determination of cosmological parameters. Looking at redshift-space distortions we 
are able to obtain maximum likelihood estimates of $\Omegam$ (Fisher, Scharf \& Lahav 
1994; Heavens \& Taylor 1995; Baker \etal 1998; Hamilton 1998 and references 
therein; Dekel 1999a,b; Hamilton, Tegmark \& Padmanabhan 2000) and the 
galaxy-galaxy power spectrum (Fisher \etal 1993; Feldman, Kaiser \& Peacock 1994; 
Lin \etal 1996; Sutherland \etal 1999; Hamilton \& Tegmark 2000).  
The dynamics of cosmic velocity flows is also another way to measure $\Omegam$ 
(Willick \etal 1997; Dekel, Burstein \& White 1997; da Costa \etal 1998; 
Kashlinsky 1998; Sigad \etal 1998; Susperregi 2001, hereafter S01) and the power 
spectrum (Strauss \& Willick 1995 and references therein; Kolatt \& Dekel 1997; 
Willick \etal 1997; Freudling \etal 1999; Zehavi \& Dekel 1999). 
In the larger picture, galaxy redshift surveys contribute modestly, alongside 
a myriad of other datasets (CMB, SN1a data, weak lensing surveys, etc) to the  
undertaking of estimating the key cosmological parameters ($\Omegam$, $\Omega_0$, 
$\Omega_B$, $\Omega_{\Lambda}$ or $\Omega_{Q}$, $h$, $n_s$, $n_t$, etc)(for a 
review of how this inventory of over a dozen parameters is currently tackled, see e.g. 
Turner 1999; Primack 2000); 
all the datasets, including galaxy surveys, can be ideally combined by 
exploiting cosmic complementarity (Eisenstein, Hu \& Tegmark 1998,1999) 
leading to demonstrably concordant, or at least not inconsistent, predictions. 

A number of techniques have been employed to maximize the parameter 
information extracted from surveys. The Fisher information matrix approach 
is one example of this (Tegmark 1997; Tegmark, Taylor \& Heavens 1997; 
Goldberg \& Strauss 1998; Taylor \& Watts 2000). This approach is sound and 
will lead to very accurate results when applied to forthcoming surveys such 
as SDSS and 2dF; the computational effort can be successfully 
minimised in those cases via data-compression techniques, e.g. via the 
Karnhunen-Lo\`{e}ve ``signal-to-noise'' eigenmodes (Vogeley \& Szalay 1996;  
Tegmark \etal 1998; Matsubara, Szalay \& Landy 2000), quadratic compression into 
high resolution powers (Tegmark 1997; Tegmark, Taylor \& Heavens 1998; Tegmark \etal 
1998; Padmanabhan, Tegmark \& Hamilton 2000), etc. There is, on 
the other hand, a case to be made for combining different 
galaxy redshift surveys and extracting information jointly 
from them, as an ensemble, prior to comparing them with different kinds 
of datasets. As a step prior to exploiting cosmic complementarity, 
it can be argued that redshift surveys are best taken into account 
as an ensemble than individually, however larger and better 
sampled is any given dataset in comparison to its predecessors. 
It is to be expected that a comparison of surveys leaves considerable 
room to manoeuvre to best rid of, or, at least compensate for, the 
errors inherent in each dataset, stochasticity, unaccounted for 
contaminants, etc. 

The goal of this paper is to set out a procedure to extract an accurate 
estimate of $\Omegam$ from galaxy redshift surveys, when the datasets 
are studied as an ensemble. The measurement of $\Omegam$ will then be the 
optimal estimate for all the surveys considered. The method will also 
enable us to estimate the matter and galaxy-galaxy power spectra. The backbone 
of the procedure is essentially the least-action principle (LAP) reconstruction 
employed in S01. That paper showed that the LAP method is efficient to  
measure $\Omegam$ from a given galaxy redshift survey and that it 
conveniently breaks the degeneracy between $\Omegam$ and the bias, 
thus permitting to measure $\Omegam$ within fairly arbitrary bias schemes. 
In this paper we consider a bias relationship that is scale-dependent in 
order to investigate the dynamics more realistically. Bias is in any case 
stochastic as well as scale-dependent (Pen 1998; Tegmark \& Peebles 1998; 
Dekel \& Lahav 1999; Tegmark \& Bromley 1999;  Taruya 2000), and the 
bias relationship is fully determined by the distribution $P(\delta_{\rm g}|\delta_{\rm m})$ 
(Dekel \& Lahav 1999; Sigad, Branchini \& Dekel 2000). The stochastic ingredient 
is largely sample-dependent, and the local scatter that is the measure of stochasticity 
in $P(\delta_{\rm g}|\delta_{\rm m})$ is a characteristic of each survey, 
its shot noise and environmental effects. The whole point of studying an ensemble 
of surveys is to minimise this effect, and therefore for simplicity we 
neglect the existence of stochasticity in the bias and we focus on its scale 
dependence.  
 
The galaxy redshift surveys that we examine in this paper are 
PSC$z$, ORS, Mark III and SFI. The LAP method is initially applied to 
each survey individually in the manner of S01 with a scale-dependent 
bias model. Hence one obtains the cosmic fields and a measurement of 
$\Omegam$ for each survey. Each survey is regarded as a constraint 
on the cosmic fields, and as they are all indeed different, the 
system that results from considering all constraints simultaneously is 
unavoidably overconstrained. Strictly speaking it is also incompatible. 
The method to find an optimal solution that is a meaningful representation of 
the ensemble entails relaxing all the constraints. The errors that this 
relaxation brings in are dealt with the least-square method, and this 
yields the best fit. The procedure gives us also an estimate of $\Omegam$, which 
is really a joint measurement for all the surveys in the ensemble. From the 
cosmic fields and the bias relationship it is then straightforward to compute 
the power spectra, higher-order correlation functions, etc. 

The article is structured as follows. A brief description of the 
LAP method is given in \S 2 and the algebra is relegated to the appendices, though 
the necessary steps to undertake the numerical implementation are set out in 
\S 2.1; the characteristics of the surveys that are relevant to this paper 
are summarized in \S 3; the procedure for finding LAP solutions for the 
ensemble of datasets is described in \S 4 for the most general case; in \S 5 
we discuss the application of the method on the four surveys in question and 
their results; finally \S 6 discusses the main results of this paper, in comparison 
with similar measurements in the literature.  

\section{LAP preliminaries}

The three ingredients of the LAP method are: (a) two boundary conditions and  
(b) the Lagrangian of the self-gravitating matter field. The two boundary conditions 
largely determine the cosmological model, as they are the endpoints in the 
evolution of our Universe. Initial homogeneity, as measured by the Sachs-Wolfe 
constraint on the CMB, suggests the first is  
\be \label{cons1}
\delta (t\to0,{\bf x})\approx0,
\ee 
where $\delta$ is the matter density constrast. The statistics of the primordial 
$\delta$ is thus unconstrained. The second boundary condition is the Universe 
as is represented in the current epoch by a galaxy redshift survey; this is characterised 
by the galaxy number-count $n({\bf s})$ ($\bf s$ denotes redshift coordinates, 
($cz$,$\theta$,$\varphi$)). This constraint is given by (S01) 
\be \label{cons2}
\rho_s ({\bf s})= x^2 {N_{\rm gals}\over V}
\biggl[{1+ g_0({\bf x})\over 1+\alpha_0({\bf x})^{\prime\prime}}\biggr], 
\ee 
where $\rho_s$ relates to the galaxy number-count via 
$\rho_s\equiv \d n({\bf s})/\d s\d {\sl\Omega}$ (where $\d {\sl\Omega}$ 
denotes a unit of solid angle), $V$ is the volume of the survey, 
$x$ is the radial comoving distance, $\alpha$ is the velocity potential 
(i.e. ${\bf v}\equiv\nabla\alpha$), the prime denotes the 
line-of-sight derivative $\d/\d x$, $g \equiv \delta_{\rm gals}$ is the 
galaxy number density contrast and the subindex $0$ denotes the present time. 
We use comoving quantities consistently throughout the paper. 
The bias model that we shall adopt is scale-dependent, such that 
\be \label{bias}
g_0({\bf k})= b(k)\delta_0({\bf k}). 
\ee
The method is used following the same procedure as that expounded in S01, with 
the addition of a scale-dependent bias. The revelant LAP equations are summarized 
in Appendix A for reference. 

\subsection{Numerical resolution}

We apply the algorithm as follows. 
\begin{itemize}
\item The dataset $\cal D$ is transformed into a $z$-space field 
$\rho_s({\bf s})$, by computing the discrete derivatives of the galaxy 
number-count $n({\bf s})$, binned to the smallest redshift spacing $\Delta z$, 
and then the result is smoothed choosing an 
appropriate smoothing length $r_s$. We shall  
only implement Gaussian smoothing, $W(k)=\exp(-k^2r_s^2/2)$. 
\item The goal is to find a best fit for the modes $\delta_y(t)$, $\alpha_y(t)$ 
(where $y\equiv rlm$ in the spherical harmonics and Bessel functions 
decomposition, following the formalism of Appendix A). A consistent 
method to achieve this is to get progressively closer to the real 
solution by trying Ansatze of increasing non-linearity in 
successive iterations. The starting point is the linear solution, which is 
approximated by inverting the relation $\delta_s\propto -\nabla^2\alpha_0$, 
where
\be  \label{delta-s}
\delta_s\equiv (s_{\rm max}/4\pi N_{\rm gals})\rho_s -1.
\ee 
For the linear $\delta$ field, we estimate 
\be \label{first-ansatz}
\delta_0\propto \delta_s({\bf x}+\hat{{\bf x}}\alpha_0^{\prime}).
\ee
In the surveys where the radial velocities are part of the observational input, in 
our case Mark III and SFI, the $v_r$ data are used to construct the first Ansatz for 
$\alpha_0^{\prime}$, which is then used to compute (\ref{first-ansatz}). 
\item The linear fields $\delta_{y,0}$,$\alpha_{y,0}$ are 
the first Ansatz that we input in the LAP system (\ref{full-sys}). We transform 
these first into the coefficients $\delta_y^{(n)}$, $\alpha_y^{(n)}$, using 
(\ref{dyn}), (\ref{ayn}). These are the coefficients of the polynomials 
of order $N$ that approximate the fields. This initial Ansatz is equivalent to 
linear theory and it permits us to solve the homogeneous system. It requires an initial 
guess of the value of $\Omegam$, as the matrix coefficients ${\cal S}^{\delta}$ 
depend on it. It is worthwhile to experiment with a ample range of values such as 
$0.1\lsim\Omegam \lsim 1.0$, to examine the impact of these variations in the 
convergence of the solutions. The solution obtained from the homogeneous system 
is least-square fitted to (\ref{cons1}),(\ref{cons2}). This in turn requires an  
bias model, and from (\ref{bias}) using Parseval's theorem (i.e. $\langle f\cdot g\rangle
=\langle f\rangle\cdot
\langle g\rangle$, where $\langle f\rangle$ denotes the Fourier transform of $f$), we have that 
$g({\bf x})=b(x)\delta({\bf x})$. In the spherical harmonics and Bessel function 
decomposition this means 
\be \label{bias-coeffs}
g(t,{\bf x})= \sum_y b_{rl}\delta_y(t) j_l(k_rx)\,Y_{lm}(\theta,\varphi).
\ee 
This assumes that the effect of bias reflects on a superposition of radial 
modes that are rotationally invariant. In this paper we have fitted $b_{rl}$ for 
$r\leq 5$ and $l\leq 6$. Our starting point Ansatz in (\ref{cons2}) is 
$b_{rl}=1.0$.   
\item The resulting $\delta_y$,$\alpha_y$ are then used to construct the RHS of 
(\ref{full-sys}), the inhomogeneous system. These quadratic terms represent 
the coupling among the normal modes of the perturbations, and their amplitude 
is a measure of the degree of non-linearity present. 
\item Successive iterations thus solve (\ref{full-sys}) and we combine this 
operation with the least-square fitting the solution to the constraints 
(\ref{cons1}),(\ref{cons2}). The procedure eventually yields the correct 
$\delta_y$, $\alpha_y$ if the errors in satisfying (\ref{cons1}), (\ref{cons2}) 
are monotonously decreasing after each iteration. We use variations around 
the solution to construct the Fisher information matrix   
\be  \label{fisher}
{\cal F}_{ij}\equiv \Big\langle {\partial^2 {\cal L}(\delta,\alpha)\over 
\partial\theta_i\,\partial\theta_j}\Big\rangle= -\int\!\d\rho_s\,
{\partial^2\over\partial\theta_i\,\partial\theta_j}\,\ln\rho_s(\delta,\alpha),
\ee
where $\theta_0\equiv \Omegam$ and $\theta_i\equiv b_{rl}$ for $i> 0$, conveniently 
ordering the $b_{rl}$ in a sequence. The likelihood function $L$, is defined 
as $L\equiv \exp(-{\cal L})$. (\ref{fisher}) enables us to compute 
the relative likelihood of estimates of the parameters, particularly $\Omegam$. 
The working assumption is that $\delta_s$ in is a random field, resulting 
from the galaxy number-count, which is modelled as a 3D Poisson 
distribution. This enables us to compute ${\cal L}$. 
The strength of the LAP method is that it does not require 
a prior, such as the CDM power spectrum, to estimate the likelihood of the parameters.
\end{itemize}  

\section{Datasets} 

\subsection{PSC$z$} 

The sample contains 14,819 galaxies, within a spherical 
region of radius $x_{\rm max}\gsim 45,000$ \km (its median is at 
$\approx 8,500$ \km and the galaxy number-count tails off  
sparsely up to $\gsim 30,000$ \km), extracted 
form the IRAS Point Source Catalogue (Saunders \etal 2000),  
following the selection procedure given by conditions 
log$(f60/f25) > 0.5$ and log$(f100/f60) < 0.75$ 
(where $f\alpha$ denotes the measured flux at $\alpha$ microns). 
The former discards all stars whilst the latter excludes 
galactic cirruses located away from the Zone of Avoidance (ZoA). 
The result retains most galaxies plus about 300 mixed contaminants, 
which are excluded on a one-to-one basis. The source catalogue contains 
17,060 sources, of which 1,593 are discarded as they are either 
within the Milky Way or repeated entries; a further 648 are rejected as 
very faint galaxies and unidentifiable sources. The resulting 6,500 are a combination 
of IRAS 1.2 Jy and QDOT galaxies plus 3,000 additional sources. The sky coverage is 
$\sim 84\%$ (the ZoA spans a non-axially symmetric 
region of $\langle |b|\rangle \lsim 5^{\circ}$). We adopt the unmasked survey, 
therefore maximising the sky coverage and a selected spherical subvolume of 
radius $x_{\rm max}\sim 15,000$ \km. 
General descriptive information and access to data are available at the PSC$z$ webpage  
{\sf http://www-astro.physics.ox.ac.uk/\~{}wjs/pscz.html}. 

\subsection{ORS}

The sample contains $8,457$ optically selected galaxies, 
of an angular size $\lsim 1.9$ arcmin, located within an 
approximately spherical region $\sim 18,000$ \km  
(Santiago \etal 1995; Hudson \etal 1995; Baker \etal 1998). 
The dataset is densely sampled within $x_{\rm max}\sim 8,000$ \km, 
which is the spherical subvolume that we shall consider, and it is sparse 
in the region $x\gsim 8,000$ \km. The sample is a compilation of optically-selected 
galaxies extracted from the Upsala General Catalogue (UGC; $\delta\geq -2.5^{\circ}$), 
European Southern Observatory (ESO; $\delta< -17.5^{\circ}$) catalogue and 
Extension to the Southern Observatory Catalogue (ESGC; covering the remaining 
region just south of the Celestial Equator. The sky coverage of these samples is 
respectively $34\%$,$19\%$ and $9\%$. ORS shares a number of 
galaxies in common with PSC$z$ though it contains a larger fraction of spirals. 
The predicted velocity fields in ORS and PSC$z$ are also in good agreement 
(Baker \etal 1998)). The sky coverage is 62\%, i.e. significantly more 
limited than in PSC$z$ (and even IRAS 1.2 Jy), due to 
the extinction in the ZoA, and it spans over 
$|b|\gsim 20^{\circ}$. In the sample utilized ORS is filled with 
823 IRAS galaxies in order to achieve $|b|\gsim 5^{\circ}$ coverage.  
A drawback of ORS is its poor uniformity, as extinction affects optically-selected 
galaxies more than it does IRAS galaxies. Further descriptive information and 
access to data are available at the URL: 
{\sf http://www.astro.princeton.edu/\~{}strauss/ors/}.

\subsection{Mark III} 

The sample contains radial velocities of $\gsim 3,400$ galaxies 
(Willick \etal 1995,1996,1997), compiled from several sets 
of S0 spirals and ellipticals. The sky coverage is the entire sky 
with the exception of the ZoA, at $|b|\gsim 20^{\circ}-30^{\circ}$. 
The data are located within a spherical region of radius 
$x_{\rm max}\sim 6,000$ \km that is well sampled, though it is 
fairly anisotropic, as it spans to $x_{\rm max}\sim 8,000$ \km~in 
some directions and is consigned to $x_{\rm min}\sim 4,000$ \km~in others. 
We consider the spherical subvolume of radius $x_{\rm max} 8,000$ \km. 
Distances are inferred via Tully-Fisher and $D_n-\sigma$ distance indicators and they 
entail an error $17-21\%$. The data must be carefully prepared 
to correct for Malmquist biases (following the procedure 
set out in Sigad \etal (1998) and used by Zaroubi, Hoffman \& Dekel (1999); also in 
a improved version by Dekel \etal (1999)). 
This results in the correction of the distances of 1,200 objects. Further 
descriptive information and access to data 
are available at the URL: {\sf http://redshift.stanford.edu/MarkIII/}. 

\subsection{SFI} 

The sample contains radial velocities of $\sim 1,300$  
Sbc-Sc galaxies, with inclination $\gsim 45^{\circ}$ north of 
$\delta< -45^{\circ}$ and galactic latitude $|b|\gsim 10^{\circ}$ 
(da Costa et al. 1996,1998; Haynes \etal 1999a,b). The data 
are distributed within an approximately spherical 
region of radius $x_{\rm max}\sim 7,000$ \km. Individual 
distances are computed in a similar manner as in Mark III, 
and the distance errors are $15-20\%$. A comparison with Mark III 
shows a good agreement (da Costa \etal 1996; ditto for the POTENT 
reconstruction analysis of Dekel \etal 1999) and it is also 
in agreement with the predicted velocity field of IRAS 1.2 Jy (da Costa et al. 
1998). 

\section{Method} 

The LAP method applied to one survey is completely determined by the 
following two constraints: 
\be \label{chi0}
\chi_0(\delta)=0,
\ee
\be \label{chi1}
\chi_{\rm survey}(\delta,\alpha)=0,
\ee
where (\ref{chi0}) is the condition of homogeneity (\ref{cons1}),  
and (\ref{chi1}) contains the data given by the survey (\ref{cons2}). 
Any additional dataset adds one constraint of the type (\ref{chi1}). 
Let us suppose we have $N$ such datasets. Obviously in practice all 
samples contain systematic and random errors and not all $N+1$ contraints  
are satisfied exactly by the same solution; 
therefore, the system $\chi_i=0$ with $i=0,1,\dots,N$ need not have 
a consistent solution $\delta$,$\alpha$. To find a solution, the 
overconstrained system must be relaxed so that the solution satisfies  
the constraints (\ref{chi1}) within the least margin of error, 
$\chi_i\approx \epsilon_i$, $(i>0)$, 
and we optimize this by computing the stationary quantity
\be \label{chi-stationary}
\tilde\delta\Big[\sum_{i=1}^N w_i\chi_i^2(\delta,\alpha)\Big]=0,
\ee
where $\tilde\delta$ denotes a variation, not a density 
contrast, and $w_i$ is the weight given to each survey in the sum. From the central 
limit theorem, we assume that $\epsilon_i$ are normally distributed about zero, 
i.e. $\langle\epsilon_i\rangle=0$ and $E(\epsilon_i^2)=\sigma_i$. We will therefore 
adopt the maximum likelihood values $w_i=\sigma_i^{-2}$, where $\sigma_i^2$ 
is the variance of the distribution $P[\chi_i]$. The likelihood function is then 
\be
L= (2\pi)^{-N/2}\,\Big(\prod_{i=1}^N \sigma_i^{-1}\Big)\,
\exp \Big(-{1\over 2}\,\sum_{i=1}^N {\chi_i^2\over \sigma_i^2}\Big).
\ee
The solution that satisfies (\ref{chi-stationary}) and the LAP 
equations (\ref{full-sys}) is the best fit to the $N$ surveys that one can find.

\subsection{Domain of application}

The first consideration is to determine the radius of the 
spherical volume to which the LAP solutions will be consigned.  
Ideally, one would wish to use the largest selected subvolume 
possible, in this case that of PSC$z$, $x_{\rm max}\sim 15,000$ \km.   
This would nonetheless give too little 
weight to the other datasets, as the constraints $\chi_i\approx 0$ given by the 
smaller surveys would have little impact in the outcome of $\delta,\alpha$ 
(the fractions of the selected volumes of ORS, Mark III and SFI, with respect to PSC$z$ are, 
respectively, 0.15, 0.06 and 0.10). On the other hand, $x_{\rm max}$ 
should not be too small either, as a large part of data from the larger samples 
may thus be left unused. An criterion to determine $x_{\rm max}^{\rm LAP}$ 
is the following. We assume that the weight each dataset ought to have in 
determining the volume of the LAP region is given by the density of 
its sampling, given by the ratio 
\be \label{sampling}
\eta_i \equiv {N^i_{\rm gals}\over V_i}\Big(\sum_{j=1}^N {N^j_{\rm gals}\over V_j}
\Big)^{-1}.
\ee
This measure of sampling is given in Table 1 for the surveys of our study. 
It shows that ORS and Mark III are the most densely sampled 
datasets and SFI is the sparsest. PSC$z$ is also 
on the whole fairly sparsely sampled. The volume of the LAP region is hence 
the weighed sum 
\be  \label{domain}
V_{\rm LAP}= \sum_{i=1}^N\eta_i V_i. 
\ee
Using the values given in Table 1, for the case at hand 
this results in $8,860$ \km. In fact this seems to be a good figure also 
for PSC$z$ as the median redshift of that catalogue is at $8,500$ \km. 
We shall adopt the round figure $x_{\rm max}^{\rm LAP}= 9,000$ \km.

\begin{table}
 \caption{Sampling factor in the surveys}
\begin{tabular}{@{}lccc}
            & $x_{\rm max}$  & $N_{\rm gals}$  &  $\eta$ \\
\hline
PSC$z$      & $15.0\times 10^3$ &  $14.8 \times 10^3$ &  0.11 \\
ORS         & $8.0 \times 10^3$ &  $8.3 \times 10^3$ & 0.41 \\
Mark III    & $6.0 \times 10^3$ &  $3.4 \times 10^3$ & 0.39 \\
SFI         & $7.0 \times 10^3$ &  $1.3 \times 10^3$ & 0.09 \\
\hline
\end{tabular}

\medskip
$x_{\rm max}$ is the radius of the selected spherical volume in each 
survey in units of \km, $N_{\rm gals}$ is the number of galaxies and 
$\eta$ is defined by (\ref{sampling}). 
\end{table}

\subsection{Solutions}

We proceed as follows. First, we apply the LAP method on each sample 
separately on their own domain of application, following the numerical 
procedure described in \S 2.1. Hence, for each survey we obtain 
$\delta_i$,$\alpha_i$ within a spherical volume of radius 
$x_{\rm max}^i$ (using the values given in Table 1). This also yields an estimate 
of $\Omegam$ and an evaluation of the bias coefficients $b_{rl}$. 
The results of the measurement of $\Omegam$ in each survey are given in the next section. 

Let us consider these solutions, $\delta_i$,$\alpha_i$, to construct 
an initial Ansatz of the overall solution that satisfies (\ref{chi-stationary}). This 
Ansatz is given by 
\be \label{ansatz1}
\delta=\sum_{i=1}^N\eta_i\delta_i,
\ee
\be  \label{ansatz2}
\alpha=\sum_{i=1}^N\eta_i\alpha_i. 
\ee
Thus we adopt the sampling factor as the best weight to average 
the fields. A few trial and error experiments show this is a good starting 
point. In order to consign this Ansatz to the domain (\ref{domain}), 
if $x_{\rm max}^{\rm LAP}<x_{\rm max}^i$ then one discards from the survey $i$ 
all the data that lie within the shell $x_{\rm max}^{\rm LAP}<x<x_{\rm max}^i$; and if 
$x_{\rm max}^{\rm LAP}>x_{\rm max}^i$ then ones makes the fields identically 
equal to zero within the range $x_{\rm max}^i<x<x_{\rm max}^{\rm LAP}$, effectively 
``padding'' that shell to match the LAP volume. In the example of this paper, 
PSC$z$ data are discarded in the range $9,000$ \km$<x<$15,000 \km, and ORS, Mark III and 
SFI are padded with zeros within the shell $x_{\rm max}^i<x<9,000$ \km.
 
The Ansatz (\ref{ansatz1}),(\ref{ansatz2}) is our starting point to compute the 
LAP solution over the domain (\ref{domain}). The numerical problem is then to 
solve (\ref{full-sys}), introducing variations in the fields $\delta$, $\alpha$ and 
in the free parameters $\Omegam$ and $b_{rl}$ to eventually pin down the 
stationary solution in (\ref{chi-stationary}). The numerical resolution is carried out 
following the procedure set out in \S 2.1, where all references to the constraints 
(\ref{cons1}), (\ref{cons2}) are substituted by (\ref{chi-stationary}).

\section{Results}

\begin{table}
 \caption{Likelihood of $\Omegam$}

\begin{tabular}{@{}cccccc}
$\Omegam$ & &$P_{{\rm PSC}z}$ & $P_{\rm ORS}$& $P_{\rm Mark III}$ & $P_{\rm SFI}$ \\
\hline
0.20       & & 0.00  & 0.00  & 0.00 & 0.00 \\
0.22       & & 0.00  & 0.00  & 0.00 & 0.00 \\
0.24       & & 0.00  & 0.02  & 0.00 & 0.00 \\
0.26      &  & 0.00  & 0.31  & 0.02 & 0.00 \\
0.28       & & 0.05  & 0.62  & 0.39 & 0.00 \\
0.30       & & 0.42  & 0.89  & 0.70 & 0.08 \\
0.32       & & 0.73  & 1.00  & 0.91 & 0.38 \\
0.34       & & 0.92  & 0.93  & 1.00 & 0.71 \\
0.36        && 1.00  & 0.74  & 0.93 & 0.94 \\
0.38       & & 0.91  & 0.33  & 0.72 & 1.00 \\
0.40       & & 0.70  & 0.06  & 0.40 & 0.91 \\
\hline
\end{tabular}

\medskip
The value of the likelihood distribution is normalized to unity at 
its maximum in each survey. We have binned the range $\Omegam=0.20-0.40$ in 
$\Delta\Omegam=0.02$ intervals and computed the 
solutions and the likelihood at each step.  
\end{table}

We apply the method described in \S 4 to the four datasets in question, PSC$z$, 
ORS, Mark III and SFI. For each survey we construct the galaxy 
number-count density field $\rho_s({\bf s})$ using a high-resolution 
Gaussian smoothing length of 500 \km. First of all, 
the LAP equations are solved for each survey following \S 2.1. 
This results in an estimate of the cosmic fields $\delta$, $\alpha$. 
We carry out a likelihood analysis to compute the value of $\Omegam$ in each case, 
and the same analysis also yields an estimate of the parameters 
$b_{rl}$. In order to do so, the underlying assumption is that the errors in the 
measurements of the fields and the parameters,  
$\Delta\delta_y^{(n)}$, $\Delta\alpha_y^{(n)}$, $\Delta\Omegam$ and $\Delta b_{rl}$,  
follow a multivariate Gaussian distribution. However, we do not assume 
that the fields themselves are Gaussian (not at the present time, though we find that 
they are Gaussian at the initial time, to great accuracy, and therefore it is safe to 
assume Gaussian initial conditions in the likelihood analysis though our formalism 
does not require us to do so). The results of $P(\Omegam)$ for each survey 
are given in Table 2. The maximum likelihood estimates for $\Omegam$ are 
$\Omegam^{{\rm PSC}z}=0.36$, $\Omegam^{\rm ORS}=0.32$, $\Omegam^{\rm Mark III}=0.34$, 
$\Omegam^{\rm SFI}=0.38$. The 1$\sigma$ errors in these measurements 
are of the order $\approx {\cal F}_{00}^{-1/2}$, which results 
in $2.1\times 10^{-2}$, $1.7\times 10^{-2}$, $1.3\times 10^{-2}$ and 
$1.9\times 10^{-2}$, respectively.

\subsection{Measurement of $\Omegam$}

The results obtained above are used to construct an initial Ansatz 
(\ref{ansatz1}), (\ref{ansatz2}) over the domain (\ref{domain}), 
that we use to solve (\ref{full-sys}) subject to (\ref{chi-stationary}). 
We follow the method set out in \S 4 and get the LAP solution 
that is the best fit to the four surveys.   
The likelihood analysis of the solution yields an estimate 
$\Omegam= 0.37\pm 0.01$ within the 1$\sigma$ level. This result is 
quite surprising in view of Table 2, due to the fact that the two surveys  
with the largest $\eta$, ORS and Mark III, seem to pin down $\Omegam$ 
at a lower value, in the range $0.32-0.34$, and yet PSC$z$ and SFI predict 
a $\Omegam$ that is closer to the final value. In this case, the optimization of 
the morphological features of the overdensity and velocity fields in the LAP 
solution with those of the four surveys is such that it favours a larger 
$\Omegam$. Indeed the surveys PSC$z$ and Mark III are closer to the final 
solution than ORS and SFI. The latter two yield the highest $\chi^2_i$ in 
(\ref{chi-stationary}), though SFI is in better agreement with the LAP solution 
than ORS and therefore $\chi^2_{\rm SFI}$ is smaller. PSC$z$ and Mark III have 
the smaller variance of the four, and therefore this causes that they have a 
greater weight in determining the cosmography and the cosmological 
parameters in the solution.

The likelihood analysis for the bias parameters yields the estimates 
shown in Table 3. We have computed 42 coefficients, truncating the indices 
at $r=5$ and $l=6$. The higher $l$ enables us to probe into the nonlinear 
scales, though a stringent limit is set by the smoothing scale of 500 \km. 
The high $l$ coefficients are greater than those in the linear range (low $l$), 
showing greater bias at smaller scales, a trend that is also present in 
all four surveys when the LAP method is applied to each one individually.

\begin{table}
 \caption{Bias parameters $b_{rl}$}

\begin{tabular}{@{}cccccccc}
           & & $r=0$ & $r=1$ & $r=2$& $r=3$ & $r=4$ & $r=5$ \\
\hline
$l=0$   & & 0.98 & 1.01  & 1.05  & 1.09 & 1.12 & 1.13\\
$l=1$   & & 1.02 & 1.08  & 1.11  & 1.12 & 1.15 & 1.18\\
$l=2$   & & 1.09 & 1.11  & 1.14  & 1.17 & 1.20 & 1.23\\
$l=3$   & & 1.16 & 1.18  & 1.20  & 1.23 & 1.26 & 1.30\\
$l=4$   & & 1.22 & 1.26  & 1.29  & 1.31 & 1.33 & 1.36\\
$l=5$   & & 1.30 & 1.32  & 1.35  & 1.39 & 1.41 & 1.44\\
$l=6$   & & 1.38 & 1.40  & 1.42  & 1.45 & 1.46 & 1.52\\
\hline
\end{tabular}

\medskip
Estimated values of $b_{rl}$ in the range $0\leq r\leq 5$, $0\leq l\leq 6$. 
The $1\sigma$ errors in these measurements are $\approx 0.01$ (in the worst  
case, that of $b_{43}$, it is $1.2\times 10^{-2}$, so this is the upper bound in 
the error bars of these estimates). 
\end{table}

\subsection{Matter power spectrum}

\begin{table*}
\begin{minipage}{110mm}
 \caption{Matter power spectrum}

\begin{tabular}{@{}ccccccc}
\hline
         $k$ & $P(k)$ &  $\Delta P(k)$  &  & $k$ & $P(k)$  &$\Delta P(k)$ \\
$(h\,{\rm Mpc}^{-1})$ & $(h^{-3}{\rm Mpc}^3)$ & & & $(h\,{\rm Mpc}^{-1})$
& $(h^{-3}{\rm Mpc}^3)$ &\\
\hline
 0.0182 & 9532 & 2266 && 0.1010 & 4630 & 363\\
 0.0210 & 10043 & 3451 && 0.1165 & 3981 & 298\\
 0.0242 & 11283 & 4016 && 0.1343 & 3374 & 274\\
 0.0279 & 12499 & 3220 && 0.1550 & 2935 & 260\\
 0.0322 & 13274 & 3061 && 0.1787 & 1879 & 184\\
 0.0372 & 13883 & 2874 && 0.2062 & 1140 & 165\\
 0.0429 & 14392 & 2249 && 0.2378 & 826 & 149\\
 0.0494 & 12529 & 1892 && 0.2743 & 576 & 121\\
 0.0570 & 10302 & 1220 && 0.3164 & 488 & 82\\
 0.0658 & 8821 & 921 && 0.3649 & 321 & 71\\
 0.0759 & 8427 & 887 && 0.4209 & 267 & 55\\
 0.0875 & 8033 & 408 && 0.4855 & 210 & 41\\
\hline
\end{tabular}

\medskip
Estimated values of $P(k)$ in the interval of wavenumbers 
$0.0182\, h\,{\rm Mpc}^{-1}\leq k \leq 0.4855\, h\,{\rm Mpc}^{-1}$. The range has 
been binned to produce 24 measurements within this interval. The errors, $\Delta P(k)$, 
are to 1$\sigma$ level in the likelihood estimates. 
\end{minipage}
\end{table*}

\begin{table*}
\begin{minipage}{110mm}
 \caption{Real-space power spectrum}

\begin{tabular}{@{}ccccccc}
\hline
 $k$ & $P_{\rm gg}(k)$ & $\Delta P(k)$ & & $k$ & $P_{\rm gg}(k)$  &$\Delta P(k)$ \\
$(h\,{\rm Mpc}^{-1})$ & $(h^{-3}{\rm Mpc}^3)$ & & & $(h\,{\rm Mpc}^{-1})$
& $(h^{-3}{\rm Mpc}^3)$ &\\
\hline
 0.0182 & 4232 & 2389 && 0.1010 & 5013 & 408\\
 0.0210 & 9721 & 3672 && 0.1165 & 4521 & 353\\
 0.0242 & 11396 & 4229 && 0.1343 & 3882 & 299\\
 0.0279 & 13102 & 3309 && 0.1550 & 3621 & 275\\
 0.0322 & 13855 & 3088 && 0.1787 & 2288 & 192\\
 0.0372 & 14935 & 2946 && 0.2062 & 1720 & 174\\
 0.0429 & 16223 & 2321 && 0.2378 & 1211 & 165\\
 0.0494 & 13725 & 1922 && 0.2743 & 1088 & 132\\
 0.0570 & 11263 & 1321 && 0.3164 & 975 & 89\\
 0.0658 & 8860 & 974 && 0.3649 & 720 & 80\\
 0.0759 & 8751 & 925 && 0.4209 & 612 & 62\\
 0.0875 & 8231 & 530 && 0.4855 & 488 & 53\\
\hline
\end{tabular}

\medskip
Estimated values of $P_{gg}$. The errors are, like in Table 4, 1$\sigma$. 
\end{minipage}
\end{table*}

Hereafter we use $\delta^{\rm LAP}$ and $\alpha^{\rm LAP}$ evaluated at the maximum 
likelihood values of the parameters, i.e. $\Omegam =0.37$ and $b_{rl}$ given in Table 3. 
We compute the Fourier modes $\delta({\bf k})$ of $\delta^{\rm LAP}$. The 
matter power spectrum $P(k)$ is then given directly by
\be 
\langle \delta({\bf k})\,\delta^*\!({\bf k^{\prime}})\rangle= (2\pi)^3 P(k)\,
\delta({\bf k}-{\bf k^{\prime}}),
\ee 
assuming that the modes $\delta({\bf k})$ are isotropic and homogeneous, 
and $\delta$ on the RHS is a Dirac delta function. Strictly speaking the 
fields ought be smooth, though we convolute the solutions 
with $W(k)$, in order to discard residual noise generated 
in the derivation of the LAP solution. The results for $P(k)$ are shown in Table 4. 
The galaxy-galaxy power spectrum is given by 
\be 
P_{gg}(k)\equiv b(k)^2 P(k),
\ee 
and the corresponding results are shown in Table 5. We have consigned the analysis 
to the range of wavenumbers $0.02 \,h\,{\rm Mpc}^{-1}\lsim k \lsim 0.49 \,h\,{\rm Mpc}^{-1}$ 
that we have binned in 24 equidistant segments in the log scale. 
The non-linear part of the power spectrum is located at 
$k\gsim 0.3 \,h\,{\rm Mpc}^{-1}$, and it is in this range that we 
expect the LAP method to yield significant information. 
The resulting velocity dispersion is estimated $\sigma_v\approx 286\pm 32$ \km 
and $\sigma^2_8=0.64\pm 0.03$. A COBE-normalized 
flat $\Lambda$CDM model, untilted and with $\Omegam=0.37$ and $\Lambda=0.63$, 
would fit the measured $P_{gg}$ for $\Gamma=0.21\pm 0.03$  
within the 1$\sigma$ error margin. The errors in Tables 4 and 5 are 1$\sigma$. In  
the case of Table 5, they are larger as they result from the product of 
the errors in $P(k)$ and $b(k)$. In addition to the 1$\sigma$ errors that are the 
result of the likelihood analysis, another source of error is due to the limited 
number of $\delta_y$ modes, which is propagated in the coordinate transformations 
$\delta_y\to\delta_{\bf k}$ and $b_{rl}\to b(k)$. These amount to $\lsim 10\%$ of the 
total error, and have been accounted for. 

Other power spectra of interest are (see e.g. Pen 1998; Tegmark \& Peebles 1998; 
Dekel \& Lahav 1999) the 
galaxy velocity power
\be 
P_{gv}(k)\equiv r(k)b(k)f P(k)
\ee
and the velocity-velocity power
\be \label{power-vv} 
P_{vv}(k)\equiv f^2P(k)
\ee
where $|r(k)|\geq 1$ is a galaxy-velocity correlation coefficient, 
$f\approx \Omegam^{0.6}$ is the linear growth rate. Typically $r\approx 1$ 
(Hamilton, Tegmark \& Padmanabhan 2000), though 
it may depart from unity in the nonlinear regime. Using the measurement of 
$\Omegam$ given in \S 5.1 and Table 4, it is straightforward to compute (\ref{power-vv}). 
The resulting errors result from the product of $\Delta P(k)$ and $\Delta\Omegam= 0.01$.

\section{Discussion}

The measurement of $\Omegam$ that we have obtained by applying the LAP method simultaneously 
on PSC$z$, ORS, Mark III and SFI is $\Omegam=0.37\pm 0.01$. This is consistent 
with the estimates that we have obtained from each survey separately (see Table 2), 
though ORS and Mark III predict a somewhat lower value, $\Omegam\approx 0.32-0.34$. 
The main factors in the determination of $\Omegam$ in our analysis are chiefly two: 
(1) the variance of the distribution $P[\chi_i]$ defined in \S 4, (2) the errors 
$\chi_i^2$, that reflect the resemblance of cosmographical features in the survey and 
the LAP soltuion. The ideal combination is a small variance and a minimal value of $\chi_i^2$, 
indicating a close resemblance between the survey and the LAP solution. The surveys 
that satisfy these requirements best are PSC$z$ and Mark III. The variance of ORS and 
SFI is not much greater than that of PSC$z$ and Mark III, so these surveys are consistent 
with the final LAP solution by an error of amplitude no greater than $6\%$.

These results indicate a higher value of $\Omegam$ than reported in S01 for 
IRAS 1.2 Jy ($\Omegam=0.30$ with a linear bias of $b=1.1$). In order to 
compare the results of this paper with other measurements reported in the 
literature, most of which are in the form of the redshift distortions parameter 
$\beta\approx \Omegam^{0.6}/b$, we need to have some information about the 
optimal value of the linear bias in those measurements. Hamilton (1998) surveys 
measurements up until mid-1997, most of which fall within the region 
$\beta\approx 0.45-0.75$. A higher measurement of $\beta=0.89\pm 0.12$ is reported 
by Sigad \etal (1998) from Mark III, and recently a much lower one of 
$\beta= 0.41_{-0.12}^{+0.13}$ by Hamilton, Tegmark \& Padmanabhan (2000) from  
PSC$z$ and a moderately low one of $\beta=0.5\pm 0.1$ by Nusser \etal (2000) 
from a comparison between ENEAR peculiar velocities and the PSC$z$ gravity field. 
Our results are consistent with $\beta$ measurements at the 
lower part of the range, as these allow a density parameter in the region 
$\sim 0.3-0.4$ for reasonable values of the linear bias $b\approx 1.0-1.3$. 
Higher values of $\beta$ require a much larger $\Omegam$ that is not 
consistent with our results. Such is the case of Zehavi \& Dekel (1999), who predict 
$\Omegam\gsim 0.6$ for both Mark III and SFI, and a value of $\Omegam\approx 0.4$ 
that would be consistent with our results has a low likelihood in their analysis. 

In Table 4 and 5 we have given the values of the matter and galaxy-galaxy power 
spectra respectively. The values of $P_{gg}$ are consistent, within the 1$\sigma$ 
errors given, with the results of Sutherland \etal (1999) and Hamilton \& Tegmark (2000) 
for PSC$z$. The range of 
wavenumbers considered by Hamilton \& Tegmark (2000) is 
considerable, $0.01\,h\,{\rm Mpc}^{-1}\lsim k \lsim 300\,h\,{\rm Mpc}^{-1}$, 
and they carry out a careful analysis of errors by computing the prewhitened power 
spectrum. Using similar results, Hamilton, Tegmark \& Padmanabhan (2000) fit 
their measured $P_{gg}$ to a COBE-normalized, untilted flat $\Lambda$CDM, $\Omegam=0.3$ and 
$\Omega_{\Lambda}=0.7$ (Eisenstein \& Hu (1998)), within a range of wavenumbers 
$0.01\,h\,{\rm Mpc}^{-1}\lsim k \lsim 1.00\,h\,{\rm Mpc}^{-1}$. Our results is 
in agreement with theirs, within the more limited range of wavenumbers that we 
have estimated, namely $0.01\,h\,{\rm Mpc}^{-1}\lsim k \lsim 0.49\,h\,{\rm Mpc}^{-1}$, 
and our analysis favours a slightly higher value of $\Omegam$ that, within the errors, 
is fairly close to $\Omegam\approx 0.4$. Furthermore, our analysis predicts a value 
of the velocity dispersion $\sigma_v=286\pm 32$ \km, slightly lower 
than one of the values adopted by Sutherland \etal (1999), 
$\sigma_v=300$ \km. The analysis of peculiar velocities in Mark III and SFI, 
by Zaroubi \etal (1997) and Freudling \etal (1999) respectively (see also 
Zehavi \& Dekel (1999)), yields a high-amplitude estimate of the power 
spectrum, $P(k=0.1\,h\,{\rm Mpc}^{-1})\,\Omegam^{1.2}=(4.5\pm 2.0)\times 10^3\,
(h^{-1}\,{\rm Mpc})^3$. Our estimates are consistent with this (from Table 
4, we have $P(k=0.1\,h\,{\rm Mpc}^{-1})\approx (4.63\pm 0.36)\times 10^3\, 
(h^{-1}\,{\rm Mpc})^3$), 
pinning down the values to greater accuracy. It is also the case that in our 
estimates of $P_{gg}(k)$ the error bars are significantly smaller than in 
Hamilton \& Tegmark (2000) and Sutherland \etal (1999).

\section*{acknowledgement}

This research has been funded by the Eusko Jaurlaritza research 
fellowship BFI99.116 and in part at EHU by research grant UPV172.310-G02/99.

\appendix

\section[]{LAP equations}

The cosmological perturbations are governed by (S01)
\be  \label{lagrangian}
{\cal L}={1\over 2}(1+\delta){\bf v}^{2} 
+\alpha\Big\{\dot\delta+{\bf \nabla}\cdot[(1+\delta){\bf v}]\Big\}   
-\phi\delta-{1\over 3}\Omegam^{-1}|{\bf \nabla}\phi|^2, 
\ee
where $\phi$ is the gravitational potential created by the 
matter density contrast $\delta$. The variational equations, namely 
$\tilde{\delta}{\cal S}/\tilde{\delta}\delta=0$ and 
$\tilde{\delta}{\cal S}/\tilde{\delta}\alpha=0$ ($\tilde{\delta}$ 
denotes a variation), yield: 
\be \label{zveloc-alpha}
{\bf v}={\bf \nabla}\alpha,
\ee
\be \label{zpoisson}
{\bf \nabla}^2\phi={3\over 2}\Omegam\delta, 
\ee
and 
 \be  \label{zbern}
\dot\delta+\nabla\cdot[(1+\delta){\bf v}]=0,
\ee
\be \label{zbern2}
\dot\alpha+{1\over 2}|{\bf \nabla}\alpha|^2+\phi=0.  
\ee
The radius of the Universe $a(t)$ is scaled out alongside $h$ with our use of 
units of km/s for the distance scale. We seek solutions for 
$\delta$,$\alpha$ by utilizing the trial fields 
\be \label{zd1}
\delta(t,{\bf x})=\sum_{y}\delta_{y}(t)j_l(k_rx)\,Y_{lm}(\theta,\varphi),
\ee
\be \label{zd2}
\alpha(t,{\bf x})=\sum_{y}\alpha_{y}(t)j_l(k_rx)\,Y_{lm}(\theta,\varphi),
\ee
where $y\equiv rlm$. By substituting 
(\ref{zd1}),(\ref{zd2}) into (\ref{zveloc-alpha}),(\ref{zpoisson}) 
we get
\be \label{v_spher}
{\bf v}(t,{\bf x}) = \sum_y \!\Big[\hat{\bf x}\, \alpha_y^{\prime}(t)\,j_l(k_rx) 
+{1\over x}\hat{\bf x}\wedge{\bf J}_y(\alpha)\Big] \, Y_{lm}(\theta,\varphi),
\ee
\be \label{phi_spher}
\phi(t,{\bf x})=-{3\over 2}\Omegam\!\!\sum_y
k_r^{-2}\,\delta_y(t)\, j_l(k_rx)\,Y_{lm}(\theta,\varphi),
\ee 
where the coefficients $\alpha_y^{\prime}$ and ${\bf J}_y(\alpha)$ are calculated 
in Appendix B. The time dependence of the coefficients $\delta_y$, $\alpha_y$ 
is modelled by a polynomial of degree $N$ that is a sum of 
Chebyshev polynomials, i.e. 
\be \label{deltay}
\delta_y(t)=\sum_{n=1}^N\!\delta_y^{(n)}T_n(t),
\ee
\be  \label{alphay}
\alpha_y(t)=\sum_{n=1}^N\!\alpha_y^{(n)}T_n(t).
\ee 
The properties of $T_n$ are summarised in Appendix C. We do not consider the $0$th order 
term in (\ref{deltay}), (\ref{alphay}) as it is a constant 
term. The polynomials (\ref{deltay}), (\ref{alphay}) that permit us to 
approximate $\delta_y$, $\alpha_y$, 
in such a way that the weighed sum of the errors in the time interval 
considered is least, are given by the coefficients
\be \label{dyn}
\delta_y^{(n)}=(\delta_y| T_n)\,(T_n|T_n)^{-1},
\ee
\be \label{ayn}
\alpha_y^{(n)}=(\alpha_y| T_n)\,(T_n|T_n)^{-1},
\ee
where $(|)$ is the discrete version of the weighed product (\ref{cross2}), 
i.e.
\be \label{discrete}
(f|g)\equiv \sum_{m=1}^M w(t_m)\,f(t_m)\,g(t_m),
\ee
where $w(t)$ is (\ref{w-weight2}) and $M$ is the number of $t_m$ divisions 
on the interval $[0,1]$. The boundary conditions (S01) 
\be \label{constraint1}
0= \sum_{n=0}^N (-1)^n\delta_y^{(n)},
\ee
\be  \label{constraint2}
\rho_s ({\bf s})=x^2 \biggl({N_{\rm gals}\over V}\biggr)
\Big[1+ g_0({\bf x})\Big]
\Big[1+\alpha^{\prime\prime}_0({\bf x})\Big]^{-1}.
\ee
$g_0$ is the present galaxy number-count density contrast and $V$ is the volume 
of the survey. The coefficients $\delta_y$ enter (\ref{constraint2}) through the 
bias relationship (\ref{bias-coeffs}), that also introduces the coefficients $b_{rl}$. 
The evaluation of the modes at the present time in (\ref{constraint2}) 
is simplified by virtue of $T_n(t_0)=1$. Substituting the trial fields  
into equations (\ref{zbern}),(\ref{zbern2}), we get 
\[
\sum_y (\dot{\delta}_y\!-\!k_r^2\alpha_y)\,j_l(k_rx)\,Y_{lm}(\Omega)
= -\!\!\sum_{yy^{\prime}}\!
\biggl\{\alpha_y^{\prime} 
\delta_{y^{\prime}}^{\prime}j_l(k_rx)\, j_{l'}(k_{r'}x)
\]
\be \label{zceqI}
+{1\over x^2}\Big[\hat{\bf x}\wedge{\bf J}_y(\delta)\Big]\!
\cdot\!\Big[\hat{\bf x}\wedge{\bf J}_{y^{\prime}}(\alpha)\Big]\biggr\}\,
Y_{lm}(\Omega)\,Y_{l'm'}(\Omega),
\ee 
and
\[
\sum_y \Big({3\over 2}\Omegam \,k_r^{-2}\,\delta_y
-\dot{\alpha}_y\Big)\,j_l(k_rx)\,Y_{lm}(\Omega)
\]
\be \label{zceqII}
= {1\over 2}\sum_{y^{\prime}}
\biggl\{\alpha_y^{\prime}\,\alpha_{y^{\prime}}^{\prime}\,
j_l(k_rx)\,j_{l'}(k_{r'}x)
\ee
\[
+{1\over x^2}\Big[\hat{\bf x}\wedge{\bf J}_y(\alpha)\Big]\!
\cdot\!\Big[\hat{\bf x}\wedge{\bf J}_{y^{\prime}}(\alpha)\Big]\biggr\}\,
Y_{lm}(\Omega)\,Y_{l'm'}(\Omega).
\]
The coefficients 
${\bf J}_y(\delta)$ are given in Appendix B, as in ${\bf J}_y(\alpha)$,  
with the substitution $\alpha\to\delta$. By integrating out all coordinates, 
it is possible, though arduous, to finally arrange all the coefficients 
$\delta_y^{(n)}$, $\alpha_y^{(n)}$ in the following inhomogeneous matrix system 
for each $y$ 
\be \label{full-sys}
\left[\ba{cc}
 {\cal C}_y^{\alpha} & {\cal C}_y^{\delta} \\
 {\cal S}_y^{\alpha} & {\cal S}_y^{\delta}
 \ea
\right]
\left[\ba{c}
{{\bf \alpha}_y^{(n)}}\\
{{\bf \delta}_y^{(n)}}
\ea
\right]=
\left[\ba{c}
{{{\cal D}^y_{y^{\prime}y^{\prime\prime}}}\,\delta_{y^{\prime}}^{(m)}\,
\alpha_{y^{\prime\prime}}^{(m^{\prime})}}\\
{{{\cal E}^y_{y^{\prime}y^{\prime\prime}}}\,\alpha_{y^{\prime}}^{(m)}\,
\alpha_{y^{\prime\prime}}^{(m^{\prime})}}
\ea
\right],
\ee
where ${\cal C}^{\delta}_{y}$, ${\cal C}^{\alpha}_{y}$, 
${\cal S}^{\delta}_{y}$, 
${\cal S}^{\alpha}_{y}$, ${\cal D}^{y}_{y'y''}$ and 
${\cal E}^{y}_{y'y''}$ are computed using the orthogonality relation of the 
Chebyshev polynomials (\ref{cross2}), and the standard orthogonality 
relations for $Y_{lm}$ and $j_l$,(\ref{orthoI}), (\ref{orthoII}). 
These terms are estimated numerically. The coefficients $\alpha_y$, $\delta_y$ in 
the column matrix on the LHS of (\ref{full-sys}) are arranged vertically in order 
of increasing $n$, from top to bottom, making a total of $2N$ entries. The submatrices 
${\cal C}$ result from the integration of the LHS of (\ref{zceqI}), and the 
submatrices ${\cal S}$ from integrating the LHS of 
(\ref{zceqII}). The RHS of (\ref{full-sys}) contains 
cross coefficients coupling different $y$ and $n$. In conclusion, we have 
a set of constraints for $\delta_y^{(n)}$, 
$\alpha_y^{(n)}$: (\ref{constraint1}), (\ref{constraint2}) and (\ref{full-sys})(as 
we have integrated out all coordinates from the dynamical equations, (\ref{full-sys}) 
is really an inhomogeneous system of constraints). 

\section[]{Radial derivatives}

The radial derivative of the velocity potential can be written as  
\be
{\d\over\d x}\alpha(t,{\bf x}) =
\sum_{y}\alpha_y^{\prime}(t)\,j_l(k_rx)\, Y_{lm}(\theta,\varphi),
\ee
where, using the equality 
\be \label{jprime}
{\d\over\d u}j_l(u)=(2l+1)^{-1}
\Big[ lj_{l-1}(u)-(l+1)j_{l+1}(u)\Big],
\ee
we have  
\be \label{alphap}
\alpha_y^{\prime}=k_r\biggl[{(l+1)\over
(2l+3)}\alpha_{r(l+1)m} 
-{l\over(2l-1)}\alpha_{r(l-1)m}\biggr].
\ee
Similarly
\[
\alpha_y^{\prime\prime}= k_r^2\biggl\{{(l+1)
\over(2l+3)}{(l+2)\over (2l+5)} \alpha_{r(l+2)m}
\]
\be
-\biggl[{(l+1)^2\over(2l+3)(2l+1)}+{l^2\over(2l-1)(2l+1)}\biggr]
\alpha_y
\ee
\[
+{l\over (2l-1)}{(l-1)\over(2l-3)}\alpha_{r(l-2)m}\biggr\}.
\]
The coefficients ${\bf J}_y(\alpha)$ in 
(\ref{v_spher}) are therefore given by 
\[
{\bf J}_y(\alpha)={\alpha(l,m+1)\over 2}\alpha_{rl(m+1)}
\,(i{\bf \hat{x}}_1-{\bf \hat{x}}_2)
\]
\be \label{jlmn}
+{\beta(l,m-1)\over 2}\alpha_{rl(m-1)}
\,(i{\bf \hat{x}}_1+{\bf \hat{x}}_2)+im\,\alpha_y\,{\bf \hat{x}}_3,
\ee
where 
\be
\alpha(l,m)=\Big[l(l+1)-m(m-1)\Big]^{1/2},
\ee
\be
\beta(l,m)=\Big[l(l+1)-m(m+1)\Big]^{1/2}. 
\ee
 
\section[]{Chebyshev polynomials}

The Chebyshev polynomials of the first kind $T_n$ 
satisfy the linear homogeneous second-order differential equation
\be
(1-t^2)\,\ddot{y}-t\dot{y}+n^2y=0,
\ee 
where the dot denotes $\d/\d t$. Following the convention of Courant 
\& Hilbert (1989), they are explicitly given 
by 
\be
T_n(t)={n\over 2}\sum_{m=0}^{[n/2]}(-1)^m{(n-m-1)!\over m!(n-2m)!}\,(2t)^{n-2m},
\ee
that can be rewritten as 
\be
T_n(\cos\theta)= \cos (n\theta). 
\ee
The polynomials satisfy the orthogonality relation  
\be \label{T-ortho}
\langle T_n|T_m\rangle\equiv 
\int_{-1}^{1}\d t\,w(t)\,T_n(t)\,T_m(t) =\delta_{nm}{\pi\over 2}(1+\delta_{n0}),
\ee
where $\delta_{nm}$ is a Kronecker delta function and 
\be \label{w-weight}
w(t)=(1-t^2)^{-1/2}. 
\ee
They  
have the following property 
\be
T_n(t)=T^*_n\Big({1+t\over 2}\Big),
\ee
and their recurrence relation is given by   
\be \label{T-recur}
2\, T_n(t)\, T_m(t) = T_{n+m}(t)+T_{n-m}(t),
\ee 
for $n\geq m$. In the asymptotic regime $t\to\infty$, the zeros of $T_n$ are 
\be \label{asym}
t_m^{(n)}\approx \cos\Big({2m-1\over 2n}\pi\Big),
\ee
where $t_1^{(n)}<t_2^{(n)}<\dots$. (\ref{T-ortho})-(\ref{asym}) 
are required in the evaluation of the numerical coefficients 
in (\ref{full-sys}).

The interval of orthogonality adopted in  
(\ref{T-ortho}) is $-1\leq t\leq 1$. In our application it is more convenient 
to scale $t$ down to the interval $[0,1]$, and we do so by transforming the polynomials 
via the change of coordinate $t=2\tilde{t}-1$. Therefore 
\be 
\tilde T_n(t)= T_n(2t-1)=\sum_{m=0}^{[n/2]}\tilde{a}_m\,t^m.
\ee
The coefficients $\tilde{a}_m$ of the new polynomials are found recursively 
from the old $a_m$ through the relations
\be
\tilde{a}_m^{(j)}=2a_m^{(j-1)}-a_{m+1}^{(j)},
\ee
for $m=n-1,n-2,\dots,j;$ $j=0,1,2,\dots,n;$ 
\be
\tilde{a}_m^{(-1)}={a_m\over 2},
\ee
for $m=0,1,2,\dots, n;$
\be 
a_m^{(j)}=2^j\,a_n,
\ee
for $j=0,1,2,\dots,n;$ and $a_m^{(m)}=\tilde{a}_m$ for $m=0,1,2,\dots,n$. Thus, 
the orthogonality relation (\ref{T-ortho}) in the new domain becomes
\be \label{cross2}
\langle \tilde{T}_n|\tilde{T}_m\rangle
\equiv \int_0^1\d t\,\tilde{w}(t)\,\tilde{T}_n(t)\,\tilde{T}_m(t) =
\delta_{nm}{\pi\over 4}(1+\delta_{n0}),
\ee
where  
\be \label{w-weight2}
\tilde{w}(t)={1\over 2}\,t^{-1/2}\,(1-t)^{-1/2}.
\ee
Throughout our calculations we use $\tilde{T}_n$ and the cross-product 
defined by (\ref{cross2}), as opposed to (\ref{T-ortho}), though 
we have omitted the tildes for simplicity.

\section[]{Orthogonality relations for $Y_{lm}$ and $j_l$}

\subsection{Spherical harmonics}

A spherical harmonic is defined as (following the Condon-Shortley 
convention)   
\be
Y_{lm}(\theta,\varphi)
=\sqrt{{2l+1\over 4\pi}{(l-|m|)!\over (l+|m|)!}}\,P_l^{|m|}(\cos\theta)\,
\e^{{\rm i}m\varphi}\,(-1)^{r},
\ee
where $r=m$ for $m\geq 0$ and $r=0$ otherwise, and the $P_l^m(x)$ are the associated 
Legendre functions, which are given by  
\be \label{pl-def}
P_l^m(x)={1\over 2^l l!}\,(1-x^2)^{m/2}{\d^{l+m}\over\d x^{l+m}}(x^2-1)^l.
\ee
Our convention follows Courant \& Hilbert (1969) and differs from Abramowitz and Stegun 
(1965) by a factor $(-1)^m$. (\ref{pl-def}) can be otherwise rewritten as 
\be 
P_l^m(x)= {(-1)^m\over \Gamma(1-l)}\,\Big({x+1\over x-1}\Big)^{m/2}\,F(-l,l+1;1-l;
{1-x\over 2}),
\ee
where $\Gamma$ is the gamma function and $F$ is the hypergeometric function. 
The associated Legendre functions satisfy the orthogonality condition
\be
\int_{-1}^1 P_l^m(x)\,P_n^m(x)\,\d x= {2\over 2l+1} {(l+m)!\over (l-m)!}\,\delta_{ln}.
\ee
The spherical harmonic satisfies the 
following property 
\be
Y_l^{-m}(\theta,\varphi)=(-1)^m\,Y_l^{m*}(\theta,\varphi), 
\ee
and its orthonormal relation is given by 
\be  \label{orthoI}
\int_0^{2\pi}\!\!\d\varphi\int_0^{\pi}\!\!\d(\cos\theta)
Y_{lm}(\theta,\varphi)Y^{*}_{l'm'}(\theta,\varphi)=\delta_{ll'}\delta_{mm'}.  
\ee
Furthermore, it satisfies the equality
\be
P_l(\cos\gamma)={4\pi\over 2l+1} \sum_{m=-l}^l Y_{lm}(\theta,\varphi)\,
Y_{lm}^*(\theta^{\prime},\varphi^{\prime}),
\ee
where the directions $(\theta,\varphi)$ and $(\theta^{\prime},\varphi^{\prime})$ 
are separated by an angle $\gamma$, and $P_l$ are the Legendre polynomials, 
defined as $P_l\equiv P_l^0$. 

\subsection{Spherical Bessel functions}

The spherical Bessel function of the first kind, $j_n$, is defined as 
\be \label{j-def}
j_l(z)= \sqrt{{\pi\over 2z}}\,J_{l+{1\over 2}}(z),
\ee
where $J_{\nu}(z)$ is the Bessel function of the first kind
\be
J_l(z)=\sum_{n=0}^{\infty}{(-1)^n\over n!(l+n)!}\Big({1\over 2}z\Big)^{l+2n},
\ee
that is one of two linearly independent solutions 
of the second-order differential equation
\be
z^2\,w^{\prime\prime}+z\,w^{\prime}+(z^2-l^2)\,w=0, 
\ee
where the prime denotes $\d/\d z$. 
One can rewrite (\ref{j-def}) in the so-called Gegenbauer's generalization 
\be
j_l(z)={1\over 2}(-{\rm i})^l\int_0^{\pi}\e^{{\rm i}z \cos\theta}\,P_l(\cos\theta)
\,\sin\theta\d\theta.
\ee
The spherical Bessel function satisfies the following orthogonality relation 
\[
\int_0^1\!\!\d x x^2 j_l(k_rx)\,j_l(k_sx)
={\pi\over 4k_r }\delta_{rs}\Big\{
J_{l+{1\over 2}}^{\prime}(k_r)^2
\]
\be \label{orthoII}
+\Big[1-\Big({l+{1/2}\over k_r}\Big)^2\Big]\,
J_{l+{1\over 2}}(k_r)^2\Big\},
\ee
where $\delta_{rs}$ is a Kronecker delta function and $k_r\neq 0$, and also a closure 
relation given by 
\be
\int_0^{\kappa_{\rm max}}\!\!\d\kappa\,\kappa^2\,j_l(\kappa x)\,j_l(\kappa x^{\prime})
={\pi\over 2x^2}\,\delta (x-x^{\prime}), 
\ee 
where $\delta$ is in this case a Dirac delta function.

\bsp

\label{lastpage}

\end{document}